
\magnification 1200
\def\ni{\noindent}
\def\cel{\centerline}
\def\Par{\par \vskip .23cm}

\def\d{{\bf d}}
\def\w{{\bf\omega}}
\def\ie{{\it i.e.\ \/}}
\def\qleft{\char'134}
\def\qright{\char'42}
\cel { \bf ON THE THERMODYNAMICS OF SIMPLE NON-ISENTROPIC} \par
\cel { \bf  PERFECT FLUIDS IN GENERAL RELATIVITY$^*$}

\Par

\baselineskip 14pt

\cel { Hernando Quevedo and Roberto A. Sussman}
\cel { Instituto de Ciencias Nucleares, UNAM}
\cel { A. P. 70-543,  M\'exico D.F.04510, \ M\'EXICO} \Par
\Par
\cel { ABSTRACT} \par

We examine the consistency of the thermodynamics of irrotational and
non-isentropic perfect fluids complying with matter conservation
by looking at the integrability conditions of the Gibbs-Duhem relation.
We show that the latter is always integrable for fluids of the following
types: (a) static, (b) isentropic (admits a barotropic
equation of state), (c) the source of a spacetime for which $r\ge 2$, where
$r$ is the dimension of the orbit of the isometry group. This consistency
scheme is tested also in two large classes of
known exact solutions for which $r< 2$, in general: perfect fluid
Szekeres solutions (classes I and II). In none
of these cases, the Gibbs-Duhem relation is integrable, in general, though
specific particular cases of Szekeres class II
(all complying with $r<2$) are identified for which the integrability of
this relation can be achieved. We show that Szekeres class I
solutions satisfy the integrability conditions only in two trivial
cases, namely the spherically symmetric limiting case and the
Friedman-Roberson-Walker (FRW) cosmology.
Explicit forms of the state variables
and equations of state linking them are given explicitly and discussed in
relation to the FRW limits of the solutions.
We show that fixing free
parameters in these solutions by a formal identification with FRW
parameters leads, in all cases examined, to
unphysical temperature evolution laws, quite unrelated to those of
their FRW limiting cosmologies. \Par
\vskip.5cm
\ni $^*$Work supported by CONACYT, M\'exico, project No. 3567--E

\baselineskip 18pt plus 2pt
\vfill\eject
\ni {\bf 1. Introduction} \par
\vskip .3cm

Numerous exact solutions of Einstein field equations with a
perfect fluid source exist in the literature which do not admit a
barotropic equation of state $p=p(\rho)$, where $p$ and $\rho$ are the pressure
and matter-energy density.  The thermodynamic properties of these simple
non-isentropic fluids have not been properly examined, as there is a sort of
consensus which regards as unphysical all exact solutions with such fluid
sources.  However, there are arguments supporting the idea that barotropic
equations of state are too restrictive$^1$, and so it is surprising to find so
few references in the literature$^{2,3,4,5,6}$ studying the properties of these
fluids. There are two related levels at which this study can be posed: (1) the
consistency of the thermodynamic equations with the field equations; (2) the
physical relevance of the state variables and equations of state linking them.
Obviously, if point (1) is not satisfied, point (2) cannot be even addressed,
also, it is quite possible to find exact solutions complying with (1) but
not with (2), that is, unphysical fluids whose thermodynamics is formally
correct.

Coll and Ferrando$^3$ have formally examined and solved the question
behind point (1) above. They derived rigurously the necessary and sufficient
conditions for the
integrability of the Gibbs-Duhem relation for perfect fluid sources complying
with matter conservation. These conditions become a criterion to verify if a
perfect fluid source of a given exact solution admits what Coll and Ferrando
denote a \qleft thermodynamic scheme\qright. However, these authors did not go
beyond point (1) above (the admissibility of their consistency criterion).
The purpose of this paper is to expand, continue and complement their work
by  applying their criterion (re-phrased in a more
intuitive and practical form) to irrotational and non-isentropic fluids, and
then to deal with point (2) mentioned  above, that is, to look at the physical
nature of the
resulting equations of state in the cases when the thermodynamic equations
are mathematically consistent. The contents of this paper are described below.

We present in section 2 a summary of the equations of the
thermodynamics of a general relativistic
perfect fluid, together with the conditions for admissibility of a
thermodynamic scheme as derived by Coll and Ferrando. An immediate result
is the fact that a thermodynamic scheme is always
admissible in the following three cases: (1) the fluid is static; (2) the
isometry group of spacetime has
orbits of dimension $r\ge 2$; (3) isentropic fluid, admitting a barotropic
equation of
state $p=p(\rho)$. Section 3 is concerned with the admissibility of a
thermodynamic scheme for irrotational,
non-isentropic perfect fluids. For these fluids, we investigate the conditions
derived by Coll and Ferrando in terms of differential forms expanded
in a coordinate basis adapted to a comoving frame. The conditions of section 3
are applied in sections 4 and 5 to the irrotational perfect fluid
generalization of Szekeres solutions
(class II$^{6-12}$ and the parabolic case of class I$^{7,8,11,13,14}$),
with vanishing 4-acceleration but
non-vanishing shear. These classes of
solutions admit
(in general) no isometries and are considered inhomogeneous and anisotropic
generalizations of FRW cosmologies. An interesting result from
sections 4-5 is the fact that none of these solutions admit a thermodynamic
scheme in general, that is, with unrestricted values of the free parameters
characterizing them. We show that for parabolic Szekeres class I solutions
the Gibbs-Duhem relation is not integrable for $r <2$. However, under
suitable restrictions of their parameters, specific
particular cases of Szekeres II solutions are found to
admit a thermodynamic scheme when $r< 2$.
For the particular cases of the solutions of sections 4-5 found to be
compatible with the thermodynamic scheme, we derive in section 6 explicit
(though not unique) expressions of all state variables: $\rho$ and $p$,
particle number density $n$, specific entropy $S$ and temperature $T$, as well
as two-parameter equations of state linking them. The latter turn out to be
difficult to interpret as there is no clue on how to fix the time dependent
free parameters of the solutions. We follow a common strategy which consists
in formally identifying these parameters with the FRW scale factor and
suitable
state variables. The resulting equations of state are totally unfamiliar and
(for most cases) unphysical. In particular, their corresponding temperature
evolution laws are unrelated to those of their FRW limit. Conclusions are
preanted and summarized in section 7.

We show in appendix A that the formulation of the thermodynamic scheme
provided by Coll and Ferrando is equivalent to that given in this paper.
In appendix B we show that
parabolic Szekeres class I solutions do not admit a Killing vector of
the form suggested by Szafron in his original paper$^{7,11}$.

\par

\vskip .3cm
\ni {\bf 2. The thermodynamic scheme.} \par
\vskip .3cm

Consider the perfect fluid momentum-energy tensor

$$T^{ab}=(\rho+p)u^au^b+ p g^{ab}\eqno(1)$$

\ni where $\rho$, $p$ and $u^a$ are the matter-energy density, pressure and
4-velocity. This tensor satisfies the conservation law $T^{ab}\ _{;b}=0$ which
implies the contracted Bianchi identities

$$\dot\rho+(\rho+p)\Theta=0\eqno(2a)$$

$$h_a^b p,_b+(\rho+p)\dot u_a=0\eqno(2b)$$

\ni where $\Theta=u^a_{;a}$, $\dot u_a=u_{a;b}u^b$
and $h_a^b=\delta_a^b+u_au^b$ are respectively the expansion, 4-acceleration
and projection tensor and $\dot\rho=u^a\rho,_a$. The thermodynamics of (1) is
essentially contained in the matter conservation law, the condition of
vanishing entropy production and the Gibbs-Duhem relation. The first two are
given by

$$(nu^a)_{;a}=0\eqno(3a)$$

$$(nSu^a)_{;a}= 0\eqno(3b)$$

\ni where $n$ is the particle number density and $S$ is the specific entropy.
Condition (3a) inserted in (3b) leads to $u^aS,_a=\dot S=0$, so that $S$ is
conserved along the fluid lines. The Gibbs-Duhem
relation can be given as the 1-form

$$\w=\d S={1\over{T}}\left[ \d \left({{\rho}\over{n}}\right)+p\d
\left({1\over{n}}\right)\right]\eqno(4)$$

\ni where $T$ is the temperature and $\d$ denotes the exterior derivative.
The necessary and sufficient conditions for the integrability of (4)

$$\w\wedge \d\w=0\qquad\hbox{necessary}\eqno(5)$$

\ni subjected to
fulfilment of the conservation laws (2) and (3), are the conditions which
Coll and Ferrando denote admissibility of a \qleft thermodynamic scheme\qright.
Using the 1-form $T\w$ instead of $\w$, these authors demonstrated that
a perfect fluid source admits a
thermodynamic scheme iff there exists a scalar function $F=F(\rho, p)$
satisfying

$$\Theta=\dot F\eqno(6)$$

\ni the necessary and sufficient conditions for (6) to hold is, in turn, the
fulfilment of the constraint

$$\left( {\dot p\d \dot\rho -\dot \rho \d\dot p}
\right)\wedge \d p\wedge \d\rho =0\eqno(7)$$

As immediate results from (6) and (7), it is easy to show that the
thermodynamic scheme is always admissible in three important cases:
(1) static fluids; (2) isometry groups of spacetime
have orbits of dimension $r\ge 2$ and (3) isentropic fluids.
In the first case,
$\dot\rho=\dot p=0$, and so (7) trivialy holds, while in the second case
there are always local coordinates in which all state variables,
and in particular the set $(p,\rho, \dot p, \dot\rho)$ are functions of
only two coordinates, say $(t,x)$. Since any one of the 1-forms associated
with these variables can be expanded as $\d p=p_{,t}\d t+ p_{,x}\d x$, etc,
the wedge product in (7) vanishes. The third case is characterized by the
 existence of a barotropic equation of state
$p=p(\rho)$, implying $\d p=p_{,\rho}\d \rho$ and $\dot p=p_{,\rho}\dot\rho$.
Hence, the wedge product in (7) also vanishes. Since a barotropic equation
of state can be shown to be equivalent to $\d S=0$ and $\dot S=0$,
the specific entropy in fluids compatible with such an equation of state
is a global constant (isentropic fluids). For more general, non-isentropic
fluids, this quantity is a different constant for different
observers comoving with the fluid. Also, a relation between $p$ and $\rho$
(and between any other pair of state variables) necessarily involves a third
state variable (two-parameter equations of state).

 From these results, it is clear that perfect fluids which might not comply
with the thermodynamic scheme are non-isentropic and sources of exact
solutions with weaker spacetime symmetries: with isometry groups of dimension
$r<2$. Since exact solutions of this type are difficult to obtain, in general,
we will deal in the following sections with the important particular case of
irrotational perfect fluids.

\par

\vskip .3cm
\ni {\bf 3. Thermodynamics of a non-isentropic irrotational perfect fluid.}
\par
\vskip .3cm

Consider an irrotational perfect fluid (vanishing vorticity tensor,
$\omega_{ab}=0$). Since the 4-velocity is hypersurface orthogonal, there are
local comoving coordinates $(t, x^i)$, such that the metric, 4-velocity,
4-acceleration and projection tensor are given by

$$ds^2=-N^2 dt^2+g_{ij}dx^i dx^j\eqno(8a)$$

$$u^a=N^{-1}\delta^a_t\qquad \dot u_a=(\log N)_{,a}\delta^a_i
\eqno(8b)$$

$$h_{ab}=g_{ij}\delta_a^i\delta_b^j\eqno(8c)$$

\ni where all the metric coefficients $N$ and $g_{ij}$ are (in general)
functions of all the coordinates $(t,x^i)$. In this representation, $\dot X
=(1/N)X_{,t}$ for all scalar functions and the
Bianchi identities and conservation laws (2) and (3) become

$$\rho_{,t}+(\rho +p)(\log\sqrt \Delta)_{,t} =0\eqno(9a)$$

$$p_{,i}+(\rho +p)(\log N)_{,i} =0\eqno(9b)$$

$$n={{f(x^i)} \over {\sqrt \Delta }}\eqno(9c)$$

$$S=S(x^i)\eqno(9d)$$

\ni where $\Delta\equiv\hbox{det}(g_{ij})$ and  $f(x^i)$ appearing in (9c)
is an arbitrary
function denoting the conserved particle number distribution. Consider
the coordinate basis
of 1-forms $(\d t,\d x^i)$ associated with the comoving frame (8);
the Gibbs-Duhem relation reads

$$\w =S_{,i}\d x^i={1 \over T}\left[ {\left( {{\rho  \over n}} \right)_{,i}
+p\left( {{1 \over n}} \right)_{,i}} \right]\d x^i\eqno(10)$$

\ni where the $t$ component of $\w$ in this coordinate basis vanishes
due to (9d). A sufficient integrability condition of (10) is given by

$$\d\w=W_{ti}{{\d t\wedge \d x^i} \over {nT}}+W_{ij}{{\d x^i\wedge \d x^j}
 \over {nT}}=0\eqno(11)$$

$$W_{ti}={{p_{[,i}n_{,t]}-n^2T_{[,t}S_{,i]}} \over n}=
\left( {\rho +p} \right)\left( {{{n_{,i}} \over n}{{T_{,t}} \over T}
-\dot u_i{{n_{,t}} \over n}} \right)-\left( {\rho _{,i}{{T_{,t}}
\over T}+p_{,t}{{n_{,i}} \over n}} \right)$$

$$W_{ij}={{p_{[,i}n_{,j]}+n^2T_{[,i}S_{,j]}} \over n}=
{{T_{[,i}\rho _{,j]}} \over T}-\left( {\rho +p} \right)
\left( {{{T_{[,i}+T\dot u_{[i}} \over T}} \right){{n_{,j]}} \over n}$$

\ni where square brackets denote antisymmetrization on the corresponding
indices. The necessary and sufficient condition (5) is given by

$$\d\w\wedge \w=X_{ijk}{{\d x^i\wedge \d x^j\wedge \d x^k}
\over {n^3T^2}}+X_{tij}{{\d t\wedge \d x^i\wedge \d x^j} \over {n^3T^2}}=0
\eqno(12)$$

$$X_{ijk}=-\rho _{[,i}p_{,j}n_{,k]}$$

$$X_{tij}=\rho _{[,t}p_{,i}n_{,j]}$$

\ni Conditions (12) are entirely equivalent to (7) provided by
Coll and Ferrando. One can obtain the latter form the former simply by using
(2) and (3) (in their forms (9)). However, (11) and (12) are more intuitive
than (6) and (7), as they directly incorporate state
variables such as $n$, $S$ and $T$, and their relations with $\rho$ and $p$.
Condition (12) is also more practical than (7), as it is easier to use $n$
and $S$ from (9c) and (9d) than to compute the set $(\rho,\dot\rho,p,\dot p)$
in exact solutions in which these quantities can be quite cumbersome.
The sufficient condition (11), not examined by Coll and Ferrando, is also
helpful, since  its fulfilment guarantees that (7) (or (12)) holds.
As shown in the following sections, it is easier to test in some cases
the admissibility of a thermodynamic scheme directly from (10) and (11).
However, one must have a solution of Einstein's equations providing
$\rho$ and $p$ in terms of the metric functions in order to
verify the admissibility of the thermodynamic scheme.
Thus we will look at this scheme in known
exact solutions which are particular cases of (8), and to do so we suggest
the following procedure: (a) solve the conditions (12) and substitute the
solution into (10), thus identifying possible (non-unique) forms for $S$ and
$T$; insert the obtained forms of $T$ and $n$ into (11) in order to verify if
further restrictions follow from the sufficient conditions.
If these conditions
hold, the equations of state linking the state variables $(\rho,p,n,S,T)$
(together with their functional relation with respect to the
metric functions) follow directly from integrating it.
We shown in the following sections, for various known exact solutions with
$r<2$ and with non-isentropic fluid
sources, that the Gibbs-Duhem relation might not be integrable and if it is
integrable, the resulting definitions of $S$ and $T$ might not be unique.\par

\vskip .3cm
\ni {\bf 4. Geodesic perfect fluid: the Szekeres class II solutions.} \par
\vskip .3cm

Consider the geodesic case $\dot u_i=0$. In the comoving frame (8), (9)
leads to $N_{,i}=0$ and $p=p(t)$. Without loss of generality, the metric
in the comoving frame is
(8a) with $N=1$, so that $u^a=\delta^a_t$ and all convective derivatives
are simply derivatives with respect to the time coordinate (\ie $\dot X=X,_t$
for all
functions $X$). The remaining conservation laws (9a), (9c) and (9d) have the
same forms as given by these equations. The components of
the 1-form associated with the Gibbs-Duhem relation (10) becomes

$$S,_i ={1\over{T}}\left( {{{\rho +p} \over n}}
 \right)_{,i} \eqno(13)$$

\ni while the integrability conditions $\d\w=0$ and $\w\wedge\d\w=0$
become the particular cases of (11) and (12) given by

$$W_{ti}={{T,_t} \over T}\left( {{{\rho +p} \over n}} \right)_{,i}-
 p,_t\left( {{1
\over n}} \right)_{,i}=0\eqno(14a)$$

$$W_{ij}=T_[,_i\rho ,_{j]}-{{\rho +p} \over
n}T_{[,i} n,_{j]}=T_{[,i}S,_{j]}=0\eqno(14b)$$

$$X_{tij}=0\Rightarrow n_{[,i}\rho_{,j]}=0\eqno(15)$$

\ni  The consistency between (13), its integrability conditions (14)-(15)
and the field equations will be tested on
two important classes of known exact solutions with an irrotational and
geodesic perfect fluid source: the perfect fluid generalization of Szekeres
solution$^{6-14}$. These solutions are divided in two subclasses: I and
II. This section is devoted to the class II solutions, while class I solutions
are investigated in the next section.

The Szekeres class II solutions have been extensively studied as
inhomogeneous generalizations of FRW cosmologies. A usual representation of
the spatial metric is given by

$$g_{ij}dx^idx^j=R^2\left[ {B+P}  \right]^2dx^2+{{R^2(dy^2+dz^2)} \over {\left[
{1+{\textstyle{k \over 4}}(y^2+z^2)} \right]^2}}\eqno(16a)$$

$$P={{{\textstyle{1 \over 2}}\left( {y^2+z^2} \right)U+V_1y+V_2z+V} \over
{1+{\textstyle{k \over 4}}(y^2+z^2)}}\eqno(16b)$$

\ni where $R(t)$, $V(x)$, $V_1(x)$, $V_2(x)$ and $U(x)$ are arbitrary
functions and $k=0,\pm 1$. The function $B=B(x,t)$ must satisfy the
following constraint

$$\ddot B+{{3\dot R} \over R}\dot B-{{k(B+{\textstyle{1 \over 2}}V)+U}
 \over {R^2}}=0\eqno(16c)$$

\ni which follows from the field equations. The state variables $p=p(t)$,
$\rho$
and $n$ corresponding to the metric (16a) are

$$p=-{{2R\ddot R+\dot R^2+k} \over {R^2}}\eqno(17a)$$

$$\rho ={{2R\dot R\dot B+3(B+P)\dot R^2+k(B+3P-V)-2U} \over
{R^2(B+P)}}\eqno(17b)$$

$$n={f \over {R^3(B+P)}}\eqno(17c)$$

\ni where (9c) has been used in order to arrive to (17c). The integrability
condition (15) reduces to a 3-dimensional \qleft vector product\qright
of the type $\nabla n\times \nabla \rho =0$, whose general solution has the
form

$$\rho=a(t) n^{b(t)}+c(t)\eqno(18)$$

\ni  where $(a,b,c)$ are arbitrary functions. It is evident that equation
(18) does not hold in general, that is for arbitrary values of the free
functions $(R, B, U, V, V_1, V_2)$ characterizing the metric (16). However,
it is easy to find particular cases of the latter complying with these
equations and not admitting isometries. Given the function $B_0(t)$ and the
constant $b_0$, the choice of free parameters

$$B=B_0(t)V,\qquad f=P,\qquad U=b_0V,\eqno(19a)$$

\ni leads to a solution of (15) in the form (18) with

$$b(t)=1\eqno(19b)$$

$$a(t)={{R[-2R\dot R\dot B_0+2b_0+k(1+2B_0)]} \over {B_0}}\eqno(19c)$$

$$c(t)={{3(\dot R^2+k)} \over {R^2}}-{{a(t)} \over {R^3}}\eqno(19d)$$

\ni where $B_0(t)$ is restricted by (16c), and so must satisfy

$$\ddot B_0+{{3\dot R} \over R}\dot B_0-{{2kB_0+k+2b_0} \over {2R^2}}
=0\eqno(20)$$

\ni However, as shown by equations (18)-(19), $B_0$ must also
comply with $a(t)\ne0$, or else, $\rho$ would loose its dependence on
$(x,y,z)$ and would become the matter-energy density of a FRW spacetime.
In order to identify $T$ and $S$, we insert (19a)-(19d) into (13) yielding

$$T(t)=c(t)R^3B_0=[3R(\dot R^2+k)-a(t)]B_0
\eqno(21a)$$

$$S(x,y,z)={V\over{P}}+S_0\eqno(21b)$$

\ni where $S_0$ is an arbitrary additive constant.  Regarding the sufficient
condition (14): the part (14b) holds by  virtue of $T=T(t)$, while the part
(14a) becomes $(c(t)+p)\dot T -T\dot p=0$, a condition which is identically
satisfied if (20) holds, this can be verified by inserting (21a) and (17a)
into (14a).

Notice that satisfying
the integrability conditions (14)-(15) leads to a non-unique form for $T$,
since the function $R$ is still arbitrary. Equation (20) is a second order
inhomogeneous ordinary differential equation, having (for every choice of
$R$) two linearly independent
solutions. The possible forms for $T$, characterized
by solutions of (20), as well as equations of state linking $\rho$, $n$ an
$p$ are discussed in section 6.

\Par

\vskip .3cm
\ni {\bf 5. Parabolic Szekeres class I solutions.} \par
\vskip .3cm

A particular case of these solutions was originally examined by Szafron$^{11}$
and generalized by Bona et al.$^{13}$, and in the
framework of a two-fluid interpretation, by Sussman$^{14}$. The spatial
metric is given in spherical coordinates $x^i=(r,\theta,\phi)$ as

$$g_{ij}dx^idx^j={ (Y'+\nu'Y)^2 \over A^2 }dr^2
+e^{2\nu} Y^2 (d\theta^2 +\sin^2\theta d\phi^2) \eqno(22a)$$

\ni where a prime denotes the derivative with respect to $r$ and the functions
$Y=Y(t,r)$ and
$\nu=\nu(r,\theta,\phi)$ are given by

$$Y=(2M)^{1/3}[v+Qw]^{2/3} \eqno(22b)$$

$$e^{-\nu }=1-\sin^2{\theta  \over 2} [1-A^2-B^2-C^2]
 + \sin\theta [B \cos\phi + C \sin\phi] \eqno(22c)$$

\ni with $A(r)$, $B(r)$, $C(r)$, $M(r)$, and $Q(r)$ being
arbitrary functions. The pressure $p=p(t)$ is determined by the otherwise
arbitrary functions $v=v(t)$ and $w=w(t)$ as

$$p=-{4\over 3} {\ddot v \over v }=-{4\over 3} {\ddot w \over w }
\eqno(23a)$$

\ni while the remaining state variables $\rho=\rho(t,x^i)$ and $n=
n(t,x^i)$  take the form

$$\rho ={4  \dot \Psi \dot \Omega   \over 3 \Psi \Omega } \qquad
n={ f\over \Psi \Omega } \eqno(23b)$$

\ni with

$$\Psi = v+wQ \quad \Omega = v\Gamma + w\Delta \quad
\Gamma =M'+3M\nu '\quad\Delta =\Gamma Q+2MQ'\eqno(23c)$$

\ni In obtaining the expression for $n$ in (23$b$) we have redefined the
arbitrary function $f(x^i)$ as $f\rightarrow 3Af/(2\sin\theta e^{2\nu})$.

As with the class II solutions discussed previously, there are no isometries
in general, for arbitrary values of the free parameters. This fact is proven
in Appendix B, thus correcting the erroneous result given by Szafron$^{7,11}$,
who reported in his original paper that a one dimensional isometry group
necessarily exists in these solutions. Special cases of higher
symmetry are obtained by specifying the free parameters, in particular a FRW
limit follows if $Y$ becomes separable as $Y=Y_1(t)Y_2(r)$.

The necessary and sufficient condition (15) for admittance of a thermodynamic
scheme, with $\rho$ and $n$ given by (23b) and after lengthy  algebraic
manipulations, becomes $(x^i = r,\theta,\phi)$ \

$$\rho_{[,i}n_{,j]} = {4(v\dot w - w\dot v)\over 3\Psi^3\Omega^3}
\left[ A_1\delta^1_{[i}f_{,j]} +A_2\delta^1_{[i}\nu'_{,j]} +
       A_3\nu'_{[i}f_{,j]} \right] =0 \eqno(24a)$$

\ni where $\delta^1_i$ represents the Kronecker symbol and

$$ A_1 = \Psi \dot\Psi \left[\left({\Delta\over \Gamma}\right)'\Gamma^2
          +6M^2Q'\nu''\right] + \Omega \dot\Omega Q' \eqno(24b)$$

$$ A_2 = -6fM\{\Psi\dot\Psi[\Gamma Q' + (MQ')'] + MQ'^2(w\Psi\dot)\}
\eqno(24c)$$

$$A_3 = -6M^2Q'\Psi\dot\Psi \eqno(24d)$$

\ni It follows from (24$a$) that the thermodynamic scheme conditions are
identically satisfied if $v\dot w - w \dot v =0$. However, this relationship
implies that $w \propto v$ and, according to (22$b$), the metric function
$Y(t,r)$ becomes separable. As mentioned above, this case corresponds
to the limiting FRW cosmology. We therefore demand that $v\dot w - w \dot v
\neq 0$ and, consequently,  the expression in squared brackets in (24$a$)
must vanish, i.e.,

$$(A_1+A_3\nu'')f_{,\eta} + (A_2 - A_3 f')\nu'_{,\eta} = 0 \eqno(25a)$$

$$A_3(\nu'_{,\theta} f_{,\phi} -\nu'_{,\phi} f_{,\theta}) =0
\eqno(25b)$$

\ni with $\eta= \theta, \ \phi$. Equation (25$b$), with $A_3 \neq 0$, can
always be satisfied by fixing the
angular dependence of the arbitrary function $f$. Even if we demand the
fulfillment of the simplifying restriction $f'=0$, (25$b$) still allows a
solution which fixes the angular dependence of $f$ and imposes a condition
on the radial dependence of $\nu$. In fact, using $(22c)$ and $f'=0$ it
follows that (25$b$) holds if $B=C$, $B'\neq 0$, and $AA'=(B_0-2B)B'$ where
$B_0$ is an arbitrary constant. Equation (25$a$) contains arbitrary functions
of the time as well as spatial coordinates which can be separated by
inserting (23$c$). Then we obtain

$$v\dot v (K_1 +K_2\Gamma^2) +(vw\dot)(K_1Q+K_2Q\Gamma + K_3) +
  w\dot w (K_1 Q^2+ K_2\Delta^2+2K_3Q) = 0 \eqno(26a)$$

\ni where

$$K_1 = \left({\Delta\over\Gamma}\right)'\Gamma^2 f_{,\eta}
   - 6M\{f[\Gamma Q' +(MQ')'] - MQ'f'\}\nu'_{,\eta}\eqno(26b)$$
$$K_2= Q'f_{,\eta}\eqno(26c)$$
$$K_3=-6M^2Q'^2f\nu'_{,\eta}\eqno(26d)$$

\ni For arbitrary time functions, (26$a$) yields a system of three algebraic
equations for the spatial coefficients. It then follows that this system
allows a solution iff

$$M^2Q'^2 K_2 = M^2Q'^3f_{,\eta} = 0 \eqno(27)$$

\ni If $M=0$ the spatial metric vanishes. Moreover, for $Q'=0$ the function
$Y(t,r)$ in (22$b$) becomes separable and the spacetime is that of the
FRW cosmological models. Therefore, the only non trivial solution of (27)
is $f_{,\eta}=0$. Hence, according to (26$b$) and $(26d)$, $\nu'_{,\eta}=0$,
i.e. $A$, $B$, and $C$ in (22$c$) must be constants and the metric (22$a$),
after a suitable coordinate transformation, becomes spherically symmetric.
Another possibility for solving (26$a$) is to require a linear dependence
between the time functions, that is $(vw\dot) \propto v\dot v $ and
$w\dot w \propto v\dot v$. However, it can easily be shown that this case
leads to $w\propto v$ and, therefore, we obtain the FRW limiting case.
The analisis presented above shows that the Szekeres class I solutions
admit a thermodynamic scheme in two special cases only: the spherically
symmetric limiting case and the FRW cosmologies.

We will now consider the spherically symmetric case which follows from
(22$c$) through the conditions $B=C=0$, $A^2=1$, so that $e^\nu = 1$.
The Gibbs--Duhem relation (13) yields

$$\alpha (t)\left[ {M'\over f} \right]'
+2\beta(t)\left[ {(MQ)' \over f} \right]'
+\gamma (t)\left[{ (MQ^2)' \over f} \right]'=TS' \eqno(28a)$$

\ni where

$$\alpha(t)=-{4 \over 3}v^2 \left({\dot v \over v}\right)^.\qquad
\beta(t)=   -{4 \over 3}v^2 \left({\dot w \over v}\right)^.\qquad
\gamma(t)=  -{4 \over 3}w^2 \left({\dot w \over w}\right)^.
\eqno(28b)$$

\ni In order to identify the temperature and entropy from (28$a$), we
must impose certain functional dependence either between the time functions
or the radial coefficients. We first consider the former case and require
that the following relationships hold: $\beta(t) = \kappa_1\alpha(t)$ and
$\gamma(t)=\kappa_2\alpha(t)$ where $\kappa_1$ and $\kappa_2$ are constants.
Inserting (28$b$) into these constraints, we obtain a set of differential
equations which may easely be integrated and yield

$$\dot w = \kappa_1 \dot v +\epsilon_1 v\qquad
\dot v = {\kappa_1\over \kappa_2} \dot w +\epsilon_2 w\eqno(29)$$

\ni where $\epsilon_1$ and $\epsilon_2$ are constants of integration.
Introducing the value of $\dot v$ into the equation for $\dot w$ and
viceversa, and differentiating the resulting equations with
respect to $t$, we get

$$\left(\kappa_2- \kappa_1^2\right)\ddot v
       - \kappa_1(\epsilon_1+\epsilon_2\kappa_2)\dot v
       - \epsilon_1\epsilon_2 \dot v = 0 \eqno(30a)$$
$$ \left(\kappa_2- \kappa_1^2\right)\ddot w
       - \kappa_1(\epsilon_1+\epsilon_2\kappa_2)\dot w
       - \epsilon_1\epsilon_2 \dot w = 0 \eqno(30b)$$

\ni Since $v$ and $w$ are related by $\ddot v/v=\ddot w/w$ [cf. (23$a$)],
(30$a$) and $(30b)$ implies that $\dot v/v=\dot w/w$, a condition that leads
to the FRW limiting case. Consequently, the functional dependence must
be imposed on the radial functions contained in (28$a$). The constraints

$$\left[ {(MQ)' \over f} \right]' = \lambda_1 \left[ {M' \over f} \right]'
\qquad {\rm and}\qquad
\left[ {(MQ^2)' \over f} \right]'=\lambda_2 \left[ {M' \over f} \right]'
\eqno(31)$$

\ni with $\lambda_1$ and $\lambda_2$ being constants, can be integrated
twice yielding two algebraic equations which imply  a relationship
between the arbitrary functions $M$ and $Q$, namely

$$ Q = {\lambda_2 M + \tau_2 F + \sigma_2\over
        \lambda_1 M + \tau_1 F + \sigma_1} \eqno(32)$$

\ni where $\tau_1$, $\tau_2$, $\sigma_1$, and $\sigma_2$ are arbitrary
constants and $F=\int f(r) dr$.

The identification of $T$ and $S$ follows by inserting (31) into (28$a$),
thus we obtain

$$T(t) = -{4\over 3}\left[v^2 \left({\dot v \over v}\right)^.
+2\lambda_1 v^2 \left({\dot w \over v}\right)^.
+ \lambda_2 w^2 \left({\dot w \over w}\right)^. \right] \eqno(33a)$$
$$S ={M'\over f} + S_0 \eqno(33b)$$

\ni where $S_0$ is an additive constant. We see from (33$a$) that the
temperature is determined by the arbitrary functions $v(t)$ and $w(t)$
together with the arbitrary constants $\lambda_1$ and $\lambda_2$ which
must be appropriately specified for any given $v(t)$ and $w(t)$ in
order to ensure the positiveness of the temperature.

\Par

\vskip .3cm
\ni {\bf 6. Equations of state and FRW limits.} \par
\vskip .3cm

\ni {\bf Szekeres class II solutions.} \par
\vskip .3cm

Given a choice of $R=R(t)$, the constraint (20) provides $B_0(t)$ (or given
a choice of the latter function, $R(t)$ can be obtained).
Since $p=p(t)$ and $T=T(t)$, the functions $(a(t), c(t))$ appearing in (19)
can always be expressed as functions of either one of the pair $(p,T)$ once
the constraint (20) has been solved. Thus, a generic formal equation of
state for these solutions follows by re-writing (18) as

$$\rho(p,n)=a(p)n+c(p)={a(p)\over{R^3(p)\left[1+B_0(p)(S-S_0)
 \right]}}+c(p)=\rho(p,S)\eqno(34a)$$

\ni where $n$ has been expressed in terms of $(p,S)$ by eliminating $S$
from (21b) into (17c) as

$$n(p,S)={1\over{R^3(p)\left[1+B_0(p)(S-S_0) \right]}}\eqno(34b)$$

\ni Similar forms of the equation of state, in the form $\rho(T,n)$ and
$n(T,S)$ can be obtained by expressing $p$ in terms of $T$. These formal
equations of state are difficult to interpret, as they depend on the choice
of one of the pair $(R, B_0)$, and there is no intrinsic way to select these
functions other than using the formal analogy between the functions $R$ and
$p$ in (16a) and (17a) with the scale factor and pressure of a perfect fluid
FRW cosmology (usually obeying a \qleft gamma law\qright equation of state
$p=(\gamma-1)\mu$). This analogy implies fixing $R(t)$ from solving the
FRW relation

$$p=(\gamma -1)\mu={{3(\gamma -1)(\dot R^2+k)} \over {R^2}}
=3(\gamma -1)\left( {{R \over {R_0}}} \right)^{-3\gamma }\eqno(35)$$

\ni where $p$ is given by (17a) and $R_0$ is an integration constant.
Several authors$^{6,9-12}$ have assumed this formal identification
of these functions with their FRW analogues, hoping
to provide a sort of \qleft physical handle\qright in dealing with Szekeres
class II solutions. The latter solutions, with these
specific parameters, comply with suitably defined asymptotically FRW limits.
However, as shown below, the thermodynamics of these fluids bear no relation
with that of their FRW limits, and so this is an example
of how nice geometric features do not always correspond to nice physics.

For Szekeres class II solutions, from equations (18) and (19), the FRW limit
follows as $a(t)\to 0$, so that $\rho\to\rho(t)=c(t)\equiv\mu$ and $p(t)$
given by (17a)
become the mass energy density and pressure of the limiting FRW cosmology.
This FRW limit also implies $S\to S_0$ (entropy density a universal constant),
and so, from (34b), we have $n\to R^{-3}$ (the FRW form of $n$)).
The condition $a(t)\to 0$, together with (20) implies

$$\eqalign{&2R\dot R\dot B_0-(2kB_0+k+2b_0)=0\cr
  &2R^2\ddot B_0+6R\dot R\dot B_0-(2kB_0+k+2b_0)=0\cr}$$

\ni which leads to the deSitter solution if $\dot B_0\ne 0$, or FRW metrics
if $\dot B_0= 0$ (for $k=\pm1$, one has $B_0=-(k-2b_0)/2k$,
while for $k=0$, $B_0$ is an arbitrary constant, but $b_0=0$). For the FRW
limit, from (21a), the temperature function
becomes $T=B_0\mu R^3$, and $\mu=3(\dot R^2+k)/R^2$, so that one
recovers the expected temperature law for a FRW cosmology with a \qleft gamma
law\qright equation of state, namely: $T\propto
R^{3(1-\gamma)}\propto \mu^{1-1/\gamma}$. However, for $\dot B_0\ne 0$ and
$a(t)\ne 0$ the temperature law one obtains is unrelated to these
values.

Consider first the \qleft parabolic\qright case $k=0$. Inserting the
FRW form

$$R(t)=R_0\left[ {1+{\textstyle{3 \over 2}}\gamma (t-t_0)}
 \right]^{3\gamma /2}\eqno(36a)$$

\ni into (20) and (21a) yields the following forms for $B_0$ and $T$

$$B_0(t)=c_1+c_2t_1^{1-9\gamma / 2}-{{2b_0R_0t_1^{3\gamma / 2}}
 \over {(3\gamma +2)(3\gamma -2)R_0^2}}\eqno(36b)$$

$$T(t)=3c_1\gamma R_0^3t_1^{3\gamma -2}-{{3c_2\gamma (9\gamma -4)R_0^3}
 \over {2t_1}}-2b_0R_0t_1^{3\gamma / 2}\eqno(36c)$$

\ni where $c_1$ and $c_2$ are integration constants of (20) and
$t_1\equiv 1+{\textstyle{3 \over 2}}\gamma (t-t_0)$. Notice that $T(t)$
not only bears no relation with the FRW temperature law, but is a wholly
unphysical temperature law (for $\gamma=1$, dust, $T$ is not constant).
The particular case $c_1=c_2=0$ of (40c) was obtained by Tiomno and Lima$^6$,
who report it as unrelated to the FRW temperature law. However, these
authors erroneously claim that the FRW temperature law can be recovered
(for non-FRW cases) by simply redefining the equation of state $\rho=\rho
(p,n)$ so that it depends on three parameters ($n,p$ and a function of
the spatial coordinates). Tiomno and Lima obtained $T$ from
the sufficient integrability condition (14a) expressed as
$\dot T/T=(\partial p/\partial \rho )_n \dot n/n$. Such integrability
condition is only valid if the equation of state is of the two-parameter
form.

Regarding the case $k=\pm1$, the integration of (20) for $R$ given by its FRW
analogous form leads to cumbersome hypergeometric
functions$^{12}$. Hence it is easier to demonstrate that a FRW temperature law
is incompatible with such forms of $R$ and with $B_0\ne\hbox{const.}$
complying with (20). Assume that $T$ in (21a) takes the FRW form $T=T_{_{FRW}}
=3b_1R(\dot R^2+k)$, where $b_1$ is an arbitrary constant, equation (21a)
becomes

$$\dot B_0+{{3\dot R^2+k} \over {2R\dot R}}B_0-{{3b_1(\dot R^2+k)+k+2b_0}
 \over {2R\dot R}}=0\eqno(37)$$

\ni eliminating $\dot R$ in (37) in terms of powers of $R$ from (35),
differentiating the result with respect to $t$ and inserting the obtained
forms of
$\ddot B_0$ and $\dot B_0$ into (20) leads, after some algebraic manipulation
to $B_0=-(k-2b_0)/2k$, the relation defining the FRW limit, and so, indicating
that a Szekeres class II solution with $k=\pm1$ with $R$ and $T$ having
FRW forms is necessarily a FRW cosmology. This means that selecting the free
parameters of Szekeres II solutions in terms of their resemblance to
FRW parameters does not lead to the right thermodynamics. Therefore, other
criteria must be used in order to select these parameters.

\Par

\ni {\bf Szekeres class I solutions.} \par

As in the previous case, the physical interpretation of the thermodynamic
variables of Szekeres' class I solutions presents serious difficulties.
Consider, for instance, the temperature law (33$a$). Inserting the value

of the pressure $(23a)$ into (33$a$), we obtain

$$ T(t) = p(v^2 + 2\lambda_1 v w+\lambda_2 w^2) +
{4 \over 3}(\dot v^2 +2\lambda_1\dot v\dot w + \lambda_2 \dot w^2)
 \eqno(38)$$

\ni  This temperature law already shows an unphysical behaviour
as it does not reduce, in general, to a constant in the limiting case of
vanishing pressure. Even in the extreme case $\lambda_1=\lambda_2=0$, the
required behaviour holds only if $\dot v= const.$ This implies an
additional condition on the function $v$ which cannot be satisfied
in general.

An equation of state relating $\rho$ and $n$ can be obtained from
(23$b$) by inserting the constraints (31) and (32), and using
the definition of the entropy as in (33$b$). Then

$$\rho = {4\over 3} n [(S-S_0)(\dot v^2 +2\lambda_1\dot v\dot w
+ \lambda_2 \dot w^2) + 2\tau_1 \dot v\dot w + \lambda_2\dot w^2]
\eqno(39)$$

\ni where the baryon number density, according to (23$b$) and (33$b$),
can be expressed as

$$n = (v+wQ)^{-1} [(S-S_0)(1+\lambda_2 w ) + \tau_2 w]^{-1}
\eqno(40)$$

\ni
According to (38) the variable $t$ can always be replaced by a
function of one of the pair $p$ and $T$ (or both of them).
Furthermore, (32) and (33) can be used to express $Q$ as a
function of the entropy $S$.
Consequently, (40) may be interpreted as an equation of state
of the form $n=n(p,S)$ and (39) relates $\rho$ with $p$ and $S$.
The physical interpretation of these equations of state remains
unclear as they explicitly depend on the choice of the arbitrary
functions $v$ and $w$. Therefore, it is necessary to fix these
functions by choosing a special case of Szekeres class I solutions.
Consider the spherically symmetric Szafron model$^{11,14}$

$$ v(t) = \left({t\over t_0}\right)^{1/\gamma} \qquad w(t) =
\left({t\over t_0}\right)^{1-1/\gamma} \qquad t_0={2R_0\over 3\gamma}
\qquad \gamma\neq 0, 2 \eqno(41)$$

\ni where $R_0$ is the constant FRW scale factor. The function $M(r)$ remains
arbitrary, and the thermodynamic pressure satisfies a ``gamma law" and is
given by

$$ p = (\gamma -1)\rho = {4(\gamma -1)\over 3(\gamma t)^2} \eqno(42)
$$

\ni The consistency of the equations of state and the definition of
temperature can be analyzed by looking at their specific behaviour
under physically reasonable assumptions. In the FRW limiting case of dust
($S\rightarrow S_0$ and $\gamma =1$), the temperature law (38) shows the
correct behaviour as $T =t_0^{-2} = const.$ while the equations of state
(39) and (40) yield $\rho =0$ and $n\propto t^{-1}$, respectively. Obviously,
this result has no relation to the evolution law expected at the FRW limit
($n=\rho^{1/3}$). For radiation ($\gamma = 4/3$) the temperature evolution
predicted by (38) becomes $T\propto t^{1/2}$, and the equations of state lead
to $\rho \propto   t^{-3}$ and $n\propto t^{-3/2}$ so that
$n\propto \rho^{1/2}$; whereas the expected FRW limiting behaviour should
be $T\propto t^{-1/2}$ and $n \propto \rho^{1/4}$. These examples indicate
that the requirement of a thermodynamic scheme for the Szafron class I
solutions  leads to an unphysical behaviour of the thermodynamic variables
and equations of state.

\vskip .3cm
\ni {\bf 7. Conclusions.} \par
\vskip .3cm

We have verified the consistency of the thermodynamic equations
(i.e. admittance of a thermodynamic scheme) for two
large class of exact solutions (classes I and II
Szekeres solutions) whose sources are
non-isentropic and irrotational perfect fluids. This work has aimed
at improving the study of this type of
solutions, as classical fluid models generalizing FRW cosmologies,
in contrast to a widespread attitude of simply disregarding
them for not admiting a barotropic equation of state.

For the particular cases of the solutions examined admitting a thermodynamic
scheme, the resulting equations of state have an ellusive interpretation, as
there is no blue print on how to select the free parameters of the solutions.
We have show that formally identifying the time dependent parameter $R$ of
these
solutions with the FRW scale factor leads to unphysical temperature evolution
laws, totally unrelated to that of their limiting FRW cosmology. The question
of how to select these parameters in a convenient matter remains unsolved,
though the adequate theoretical framework to carry this task has been
presented in this paper.

It has been very interesting to find that these solutions
are not compatible, in general, with a thermodynamic scheme. This
fact seems to disqualify these solutions as classical fluids of physical
interest. However, these exact solutions can still be useful if they are
examined under a less restrictive framework than that of the simple
perfect fluid. In this context, both class I and II Szekeres solutions
heve been re-interpreted as mixtures of inhomogeneous dust and a homogeneous
perfect fluid$^{12,14}$, subjected to adiabatic interaction.
In all these cases, the thermodinamics is totally different
(and much less restrictive) than that presented here.

\Par

\vskip .3cm
\ni {\bf Acknowledgements} \par
\vskip .3cm
We acknowledge fruitful discussions with H.\ Koshko and N.\ Pochol. One of us
(RAS) thanks particularly F.D.\ Chichu for critical comments on the
manuscript. \Par

\Par

\vskip.5cm
\ni {\bf Appendix A. Thermodynamic scheme conditions}
\vskip .3cm

In this appendix we explicitly derive the condition for the existence of a
thermodynamic scheme as given by Coll and Ferrando$^3$, and show its
equivalence to the condition $\omega \wedge {\bf d} \omega = 0$.

Consider a perfect fluid with mass--energy density $\rho$, pressure $p$
and baryon number density $n$, satisfying the contracted Bianchi identities
(2). If no other conditions are specified, (2) describe an open system.
However, when an equation of state is given, the closure of the system is
obtained.  The existence of a thermodynamic scheme for the fluid is therefore
closely related to the existence condition of an equation of state. To derive
this condition one may use only the matter conservation law $(3a)$. (Condition
$(3b)$ cannot be used since it involves the entropy $S$ which can be defined
only if the fluid accepts a thermodynamic scheme.)

 From the matter conservation law we define the function $F$ by
$$ \dot F = \Theta = -{\dot n\over n} \eqno(A2)$$
and consider the one--form
$$ \Gamma = {\bf d}\rho + (\rho + p) {\bf d} F \ . \eqno(A3)$$
If $\Gamma$ is completely integrable, we can define the entropy $S$ as
${\bf d}S =\Gamma/T$, where $T$ is the integral factor which can be associated
with the absolute temperature. Now we calculate the condition for the complete
integrability of $\Gamma$ which according to Frobenius' theorem$^{15}$ is
equivalent to the vanishing of the three--form $\Gamma \wedge {\bf d} \Gamma$.
 Clearly, this condition will not be satisfied
in general unless we demand certain functional dependence for $F$.
Assuming that (A2) allows a solution of the form $F=F(\rho, p)$, it follows
that
$${\bf d} F = F_{,\rho}  {\bf d}\rho + F_{,p} {\bf d} p
\qquad {\rm and} \qquad
\dot F = F_{,\rho} \dot  \rho
             + F_{,p} \dot  p \  \eqno(A4)$$
where $F_{,\rho} =\partial F/ \partial \rho$ and $F_{,p} =
\partial F/\partial p $.
Accordingly, (A3) and the Bianchi identity (2a) may be written as
$$\Gamma = [1 + (\rho+p)F_{,\rho}]{\bf d}\rho + (\rho + p) F_{,p} {\bf d} p
\eqno(A5) $$

$$[1 + (\rho+p)F_{,\rho}]\dot \rho + (\rho + p) F_{,p} \dot p  = 0
\eqno(A6)$$
respectively, and lead to
$$\Gamma = (\rho + p)F_{,p}
          \left( {\bf d} p - {\dot p\over \dot \rho} {\bf d}\rho\right) \ .
\eqno(A7)$$
It follows from the last equation that the integrability condition for
$\Gamma$ is equivalent to
$$\Gamma \wedge {\bf d}\Gamma = -(\rho + p) F_{,p} \dot\rho^{-2}
(\dot p {\bf d}\dot \rho -\dot \rho {\bf d}\dot p ) \wedge {\bf d} p
\wedge {\bf d } \rho = 0 \ . \eqno(A8)$$
The last expression leads to the condition (7) obtained by Coll and

Ferrando.

Consider now the Gibbs--Duhem relationship as given in (4), that is
$\omega=\omega(\rho, p ,n , T)$. Since $\omega$ is a 1--form, the condition
for its complete integrability, $\omega\wedge {\bf d}\omega =0$, will be
satisfied if $\omega$ can be defined in a two--dimensional space with
thermodynamic coordinates, say, $\rho$ and $p$. This means that an equation
of state must exist such that $n=n(\rho, p)$ and $T=T(\rho, p)$. Consequently,
the Gibbs--Duhem relationship becomes
$$\omega = {1\over nT} \left\{ \left[1-(\rho+p) {n_{,\rho}\over n}\right]
{\bf d} \rho - (\rho + p){n_{,p}\over n}{\bf d} p \right\} \eqno(A9)$$
where $n_{,p}=\partial n/\partial p$, etc. Introducing the Bianchi identity
$(2a)$ [with $n=n(\rho, p)]$ into (A9) yields
$$\omega = (\rho + p) {n_{,p} \over n^2 T} \left( {\dot p\over \dot \rho}
{\bf d} \rho - {\bf d} p\right) \ . \eqno(A10)$$
It is now easy to see that the integrability condition $\omega \wedge
{\bf d} \omega =0 $ is equivalent up to a constant factor to (A8).

\Par

\vskip.5cm
\ni {\bf Appendix B. Parabolic Szekeres class I solutions and isometry
groups}
\vskip .3cm

In his original paper$^{11}$, Szafron introduced the particular class of
parabolic Szekeres class I solutions that has become known in the literature
as the \qleft Szafron models\qright. He claimed that the latter admit at
least a
one parameter isometry group, and provide specific forms for the components
of this Killing vector. However, Szafron's claim, stated
again by Kramer et al$^7$, is false. We prove in this appendix
that Szekeres class I solutions admit no Killing vector of the
type suggested by Szafron.

In section 5 we have use spherical coordinates to investigate the
Szekeres class I solutions; however, for the investigation of their
isometries it is convenient to introduce the original coordinates
used by Szafron$^{11}$ and  Bona et al.$^{13}$. The spatial metric,
save changes in notation, is given as

$$g_{ij}dx^idx^j=P^2\left[ {\left( {{{B^{2/ 3}} \over P}} \right)'}
 \right]^2dx^2+{{B^{4/ 3}} \over {P^2}}(dy^2+dz^2)\eqno(B1)$$

\ni where a prime denotes derivative wrt $x$, the function
$P(x,y,z)$ is given by

$$P={\textstyle{1 \over 2}}U(x)(y^2+z^2)+V_1(x)y+V_2(x)+V(x)\eqno(B2)$$

\ni where the functions $U(x)$, $V_1(x)$, $V_2(x)$ and $V(x)$ restricted by
the condition: $UV-V_1^2-V_2^2-{\textstyle{1 \over 2}}=0$, while $B(t,x)$ is
determined by the field equation

$$\ddot B+{\textstyle{3 \over 4}}p(t)B=0\eqno(B3)$$

\ni which can be considered as linear second order differenttial equation,
and so $B$ has the generic form $B=M(x)v(t)+Q(x)w(t)$
identifying the two arbitrary integration constants as the arbitrary
functions $M(x)$ and $Q(x)$. The two linearly independent solutions
of (B3), $v(t)$ and $w(t)$, are related to the pressure by

$$ p(t)=-{{4\ddot v} \over {3v}}=-{{4\ddot w} \over {3w}}\eqno(B4)$$

Adapted to the coordinates and notation of (B1), the generic form of the
Killing vector provided by Szafron is

$$K=Y(x,y,z)\partial_y+Z(x,y,z)\partial_z\eqno(B5)$$

\ni Computing the Killing equation for the metric (B1) and vector field (B5)
leads immediately to the conditions

$$Y_{,x}=Z_{,x}=0\eqno(B6)$$

$$Y_{,z}+Z_{,y}=0\eqno(B7)$$

$$Z_{,z}-Y_{,y}=0\eqno(B8)$$

$$\left( {{{P_{,x}} \over P}} \right)_{,y}Y+\left( {{{P_{,x}} \over P}}
 \right)_{,z}Z=0\eqno(B9)$$

\ni Although Szafron's result refers to a particular case of (B1),
condition (B6) is sufficient to prove that this result is wrong: the
components of the vector fields he provided cannot be components of a
Killing vector of (B1). This is so, because Szafron's particular case
follows by giving specific forms to the functions
$v(t)$ and $w(t)$ in (B4) and these functions are not involved in
the calculation which leads to (B6--B9). However, even assuming $Y=Y(y,z)$
and $Z=Z(y,z)$, (B5) is not an isometry of the metric (B1).
This can be proven by inserting (B2) into (B9) so that the resulting
equation

$$(yY+zZ)\left({U\over P}\right)' + y \left({V_1\over P}\right)'=0 $$

\ni implies a functional dependence between the functions $U$ and $V_1$
of the form $(V_1/P)' = G(y,z)(U/P)'$, where $G(y,z)$ is an arbitrary function
of its arguments. Obviously, the latter condition cannot hold for an
unrestricted form of $P$.

\vskip .8cm
\ni {\bf References} \par
\vskip .3cm

\ni $^1$ C.B. Collins, {\it {J. Math. Phys.}}, {\bf{26}}, 2009, (1985). See
also: C.B. Collins, {\it{Can. J. Phys.}}, {\bf{64}}, 191, (1986).
\Par

\ni $^2$ A. Lichnerowicz, (1967), {\it{Relativistic Hydrodynamics and
Magnetohydrodynamics.}}, (Benjamin, New York). \Par

\ni $^3$ B. Coll and J.J. Ferrando, {\it {J. Math. Phys.} }, {\bf 26},
(1985), 1583.\Par

\ni $^4$ C. Bona and B. Coll, {\it Gen. Rel. Grav. }, {\bf 20},
(1988), 297.\Par

\ni $^5$  R.A. Sussman, {\it J. Math. Phys.}, {\bf 29}, 945, (1988). See
sections VI-VII. \Par

\ni $^6$ J. Tiomno and J.A.S. Lima, {\it Class. Quantum. Grav.}, {\bf 6},
L93, (1989).\Par

\ni $^7$ D. Kramer, H. Stephani, M.A.H. MacCallum, and E. Herlt,
{\it Exact Solutions of Einstein's Field Equations},
(Cambridge University Press, Cambridge, UK, 1980).
On perfect fluid Szekeres solutions, see
section 29.3, pp 316. Spherically symmetric particular cases of the latter are
discussed in paragraph 14.2.4, pag. 172.

\Par

\ni $^8$ A. Krasinski,
{\it Physics in an Inhomomogeneous Universe, (a review)},
(Polish Research Committe and Department of Applied Mathematics, University
of Cape Town, South Africa). 1993.\Par

\ni $^9$ N. Tomimura, {\it Nuovo Cimento B}, {\bf{42}}, 1, (1976). See also:
W.B. Bonnor and N. Tomimura, {\it Mon, Not. Roy. Astr. Soc.},
{\bf{175}}, 85, (1976).
\Par

\ni $^{10}$ D.A. Szafron and J. Wainwright, {\it J. Math. Phys}, {\bf 18},
1668,
(1977).\Par

\ni $^{11}$ D.A. Szafron, {\it J. Math. Phys}, {\bf 18}, 1673,
(1977).\Par

\ni $^{12}$ J. Tiomno and J.A.S. Lima, {\it Gen. Rel. Gravit.}, {\bf 20},
1019, (1988).\Par

\ni $^{13}$ C. Bona, J. Stela and P. Palou, {\it J. Math. Phys}, {\bf 28}, 654,
(1987).\Par

\ni $^{14}$ R.A. Sussman, {\it Class. Quantum. Grav.}, {\bf 9},
1891, (1992).\Par

\ni $^{15}$ See, for example, Choquet--Bruhat Y, DeWitt--Morette C, and
Dillard--Bleick M 1982 {\it Analisis, Manifolds and Physics} (Amsterdam:
North--Holland Publishing) pp. 245 ff.\Par

\bye